\documentclass{llncs}

\usepackage{todonotes}
\usepackage{xspace}
\usepackage{algorithm, algcompatible}
\usepackage{url}
\usepackage{caption}
\usepackage{subcaption}
\newtheorem{observation}{Observation}

\newcommand{\firstphase}{safe phase\xspace}
\newcommand{\secondphase}{unsafe phase\xspace}
\newcommand{\Firstphase}{Safe phase\xspace}
\newcommand{\Secondphase}{Unsafe phase\xspace}
\newcommand{\homo}{Find\_Homogeneous\_Frames\xspace}
\newcommand{\single}{Find\_Single\_Option\_Voters\xspace}
\newcommand{\LE}{Likelihood\_Estimation\xspace}

\newcommand{\resultssubfigure}[9]{%
  \begin{figure}
    \centering
    \begin{subfigure}{.5\textwidth}
      \centering
      \includegraphics[scale=0.25]{#1_success_by_#2_for_#3}
    \label{fig:sub1}
    \end{subfigure}%
    \begin{subfigure}{.5\textwidth}
      \centering
      \includegraphics[scale=0.25]{#6_success_by_#7_for_#8}
    \label{fig:sub2}
    \end{subfigure}
    \caption{%
      Results for the #1 phase,
      showing the success rate as a function of the #5 (left) and the #9 (right),
      when extracting each voter's #4.
    }
    \label{figure:#1_success_by_#2_for_#3}
  \end{figure}
}

\begin{document}

\pagestyle{plain}

\title{Breaching the Privacy of Israel's Paper Ballot Voting System}

\author{Tomer Ashur\inst{1} \and Orr Dunkelman\inst{2} \and Nimrod Talmon\inst{3}}
\institute{ESAT/COSIC, KU Leuven and iMinds, Leuven, Belgium
 \and University of Haifa, Haifa, Israel 
 \and Weizmann Institute of Science, Rehovot, Israel\\
 {\tt [tomer.ashur] {@} esat.kuleuven.be}\\
 {\tt [orrd] {@} cs.haifa.ac.il}\\
 {\tt [nimrodtalmon77] {@} gmail.com}}

\maketitle

\begin{abstract}
An election is a process through which citizens in liberal democracies select their governing bodies,
usually through voting.
For elections to be truly honest, people must be able to vote freely without being subject to coercion;
that is why voting is usually done in a private manner.
In this paper we analyze the security offered by a paper-ballot voting system that is used in Israel, as well as several other countries around the world. 
we provide an algorithm which,
based on publicly-available information,
breaks the privacy of the voters participating in such elections.
Simulations based on real data collected in Israel show that our algorithm performs well,
and can correctly recover the vote of up to 96\% of the voters. 
\end{abstract}

\setcounter{footnote}{0}
\section{Introduction}
\label{sec:intro}

One of the fundamental mechanisms that allow for democracy is the notion of free elections.
In free elections,
eligible voters express their opinions on important matters via voting.
In liberal democracies,
periodical elections (which we refer to as ``election cycles'') are held for electing the members of the governing bodies.
For people to freely express their opinions
(that is, without being coerced to external pressure),
voting is usually done in a private manner.
In other words,
the elections allow voters to maintain their privacy regarding their specific vote within a large anonymity set.

One can learn about the importance of secrecy in election processes from the
\emph{Declaration on Criteria for Free and Fair Elections},
published by the \emph{Inter-Parliamentary Union} in 1994,\footnote{%
  The Inter-Parliamentary Union (IPU) is an international organization of 162 state parliaments and 10 regional parliaments.
  This union,
  which was established in 1889,
  has a permanent observer status at the United Nations
  and
  general consultative status with the Economic and Social Council.}
  and which states~\cite{IPU}:
  \begin{quote}
    ``2. Voting and Elections Rights:\\
      (7) The right to vote in \underline{secret} is absolute and shall not be restricted in any manner whatsoever.''
  \end{quote}
Similarly in spirit,
the state of Israel have recognized the importance of secret voting and determined in its
\emph{Basic Law: The Knesset}\footnote{%
  The \emph{Knesset} is the name of the Israeli parliament.}~\cite{BasicLawKnesset},
in Section 4 that:
\begin{quote}
  ``The Knesset shall be elected by general,
  national,
  direct,
  equal,
  \underline{secret},
  and proportional elections,
  in accordance with the Knesset Elections Law."
\end{quote}
In this paper,
we demonstrate that only a few observations
are required to breach the privacy of the voters in the Israeli general elections.
Our attack uses only the following information:
  (1) the results of the elections per ballot box
  (which are published at the end of the election cycle by the general elections committee);
  (2) the time of vote for each voter
  (which is collected by the various political parties);
  and (3) a periodical count of the ballots left in the tray
  (which can be collected by the members of the local elections committee who are continuously manning the ballot box).
It turns out that,
by collecting the above information over several election cycles and using it to intersect the anonymity sets,
it is possible to recover most votes.

In what follows we report on simulations performed on real data from the 2013 Israeli general elections.
We consider variable number of election cycles which the adversary is acting upon
and consider different time intervals by which the adversary is able to count the ballots left in the tray.
We mention that an attack does not have to be global, and that the adversary can focus on specific polling stations that are of interest. 

We do use some assumptions in our simulations.
First,
we assume that an adversary can periodically count the ballots;
we elaborate on this assumption in Section~\ref{ssec:tdcp}.
Second,
in the specific simulations reported here,
we assume that voters do not switch parties between election cycles;
while this assumption is not true for all voters,
it is true for most of them (as is apparent by studying recent election surveys~\cite{2009,2013}).
While this assumption somewhat weakens the results,
it is being used in the absence of sufficient real-world data about specific voters. 
We further discuss our assumptions in Section~\ref{sec:discussion}.

Expectedly, the success of our attack increases with the number of election cycles considered
and decreases (though not dramatically) when the frequency of the count is reduced.
Our simulations demonstrate that,
for example,
with only three election cycles, it is sufficient to count the ballots once in half an hour,
to recover as much as 63\% of the voters.
Moreover,
it turns out that we can correctly recover almost all votes,
reaching 100\% success in most polling stations,
and reaching 93\% on the average,
using six election cycles and counting once in half an hour.
Further,
by counting only once in an hour,
this number remains as high as 69\%.

\subsection{Related Work}

We briefly discuss several definitions for privacy in elections.
Then,
we show how the Israeli election system can be modeled as a timed-mix,
and mention several known attacks on mixes.
Our attack, described in Section~\ref{sec:attack}, is different from these attacks,
mainly since we use a significantly smaller number of observations.

Much of the discussion around e-voting systems evolves around their security. However, the security is hardly ever compared to the alternative system ``that was always used''. Interestingly, although the underlying crypto is often well understood by specialists, e-voting systems are perceived as insecure by the layman, including decision makers. In this paper we use cryptographic tools to study the behavior of a paper-based system, allowing to compare them on the same field. We believe that adopting ideas from computer science and cryptography to verify desirable properties of real-world paper-based elections is an interesting research direction.

\bigskip\noindent\textbf{Privacy in Elections}\quad
There are several definitions for privacy in elections,
most of which borrow ideas from differential privacy.
In short,
a voting system is said to preserve privacy if
it is impossible to distinguish between two scenarios,
differentiated by the behavior of several voters;
the idea is that,
if such events are indistinguishable,
then an adversary cannot infer which of them occurred in reality.
We mention several papers~\cite{bernhard2012measuring,delaune2010verifying,kusters2010game,kusters2011verifiability} in this context.
In this paper we simply quantify the number of voters whose vote we could correctly de-anonymize.
We view our definition as being more natural,
and,
contrasted with the available definitions---which are specifically tailored for e-voting,
more suited to the context of the current paper.

\bigskip\noindent\textbf{Attacks on Mixes}\quad
Mixes are widely used to model private communications.
Proposed by Chaum in 1981~\cite{DBLP:journals/cacm/Chaum81}, a mix 
is a means for delivering messages anonymously between senders and receivers.
Communication in a mix is split into rounds,
such that in each round $n$ senders send messages which are then sent to $n$ receivers in an arbitrary or random order.

Each ballot box in the Israeli voting system can be modeled as a certain kind of a mix,
namely a timed-mix. In such a mix, a buffer of messages is mixed once in each time period.
The set of voters in each polling station corresponds to the set of senders,
while the set of parties contesting in an election corresponds to the set of receivers.
There are various known attacks on mixes~\cite{DBLP:journals/ieeesp/AgrawalK03,DBLP:conf/sec/Danezis03,DBLP:conf/ih/KedoganAP02,DBLP:conf/ih/KesdoganP04,DBLP:conf/pet/TroncosoGPV08} and we refer the interested reader to a recent survey~\cite{lu2015towards}.

Most of the above-mentioned papers de-anonymize single receivers and assume either a uniform distribution of the other receivers
or try to approximate that distribution.
In our case,
the overall tally is given,
and we aim to de-anonymize the whole electorate.

\subsection{Paper Organization}

The paper is organized as follows:
  Section~\ref{sec:dotes} gives a brief description of the Israeli voting system.
  Section~\ref{sec:attack} describes our attack.
  In Section~\ref{sec:veal},
  we evaluate our attack through simulations and discuss its tightness.
  In Section~\ref{sec:discussion},
  we discuss some of the limitations of the attack,
  suggest ways to overcome these limitations,
  discuss possible countermeasures,
  and present future research directions.
  We conclude the paper in Section~\ref{sec:conc}.

\section{The Israeli Voting System}
\label{sec:dotes}

The Israeli voting system is described in the \emph{Knesset Elections law - 1969} \cite{KnessetElectionsLaw}.
In a nutshell,
every eligible citizen is assigned to a polling station.
In order to vote,
each voter arrives to her assigned polling station and identifies herself to the local elections committee.
The committee then crosses the voter's name from the list of assigned voters,
and hands her a special envelope.

\begin{figure}
\centering
\includegraphics[angle=0,origin=c,width=0.3\textwidth]{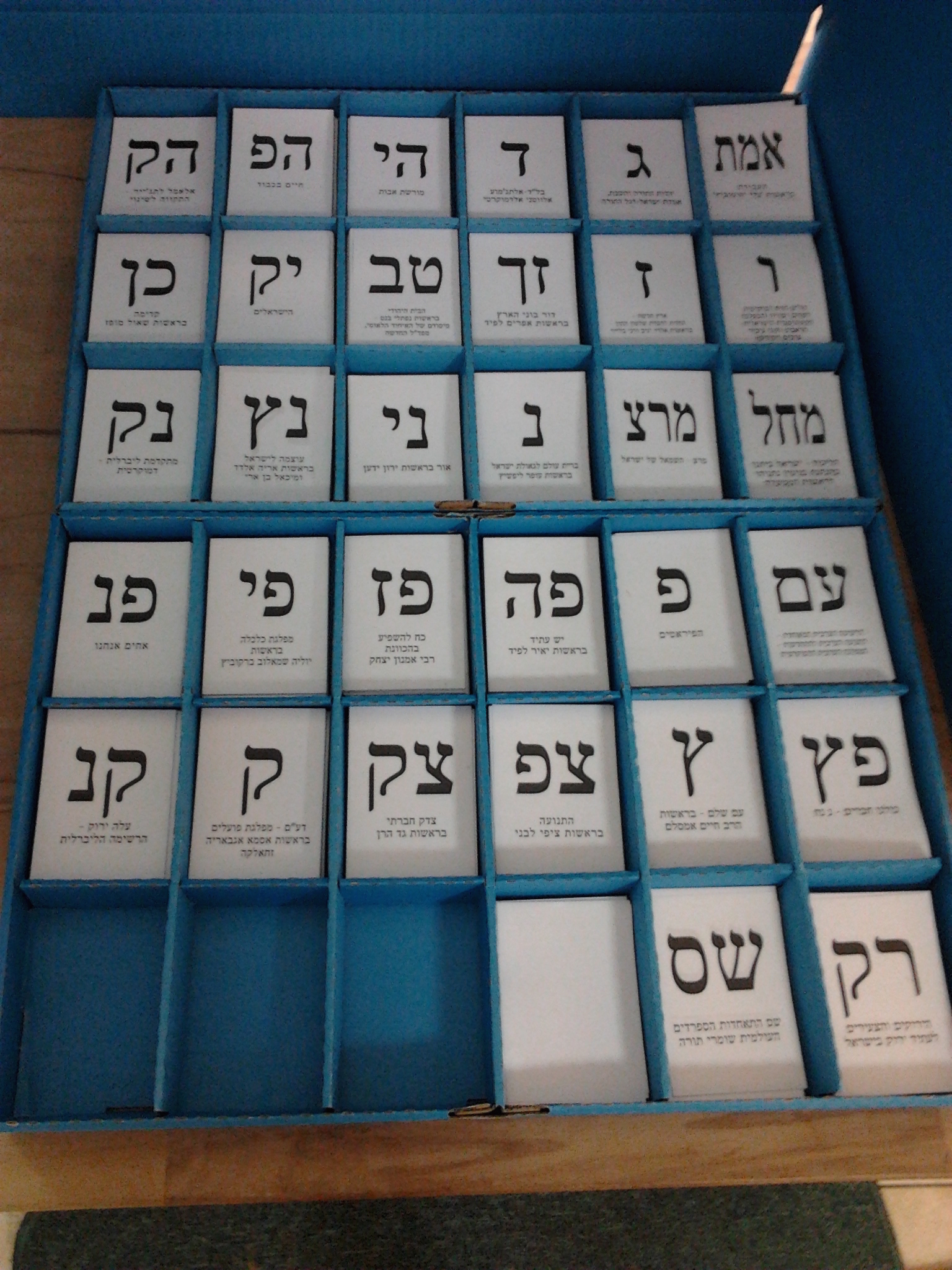}
\caption{An example of the tray for the 2013 elections.}
\label{fig:ballottray}
\end{figure}

The voter walks behind a curtain and chooses a ballot
(a piece of paper with the name of her selected party on it)
from a tray,
representing her preferred party.
The tray
(which can be viewed in Figure~\ref{fig:ballottray})
includes a stack of ballots for each candidate party
(34 parties contested in the 2013 elections).
The voter puts the ballot into the envelope,
seals,
and casts it into the ballot box,
where it mixes with all the other envelopes.
The members of the local elections committee are all,
except for the chairperson,
appointed by the political parties. As part of their role these representatives periodically check behind the curtain that all ballots are available to voters.
Another informal role of the committee members is to send the time of vote of every voter
to the parties,
so that the parties can stimulate their support base who did not show up yet,
for example, via phone calls or SMS.

At the end of the elections day,
the local elections committee breaks the ballot box's seal,
opens it,
extracts the ballots from each envelope,
and counts them.\footnote{We stress that the count is done locally, and the votes of each ballot box are not mixed with other boxes.}
The results of the tally are then sent to the general elections committee,
which aggregates and publishes the results
(including per-ballot-box statistics).
The key observation in this research is the following.

\begin{observation}
  The size of the stack of leftover ballots ``echos'' the choices made by previous voters.
\end{observation}

  For example,
  if 300 ballots are placed in the tray for each of the parties at the beginning of the election day,
  and 20 are missing from one stack after 20 voters have voted
  (and no other ballots are missing),
  then an observer can conclude that all of them voted for the party represented by this stack.

\section{The Attack}
\label{sec:attack}

In this section,
we describe our attack,
whose goal is to reveal the votes in Israel's general elections.

\subsection{Collecting Observations}
\label{ssec:tdcp}

The adversary collects observations over several election cycles $u = [1, \ldots, U]$.
For each election cycle,
in order to collect the required observations,
the adversary counts all the ballots in the tray at the beginning of the elections day.
We define this count to be in time $t = 0$.

Then,
the adversary starts counting the ballots in the tray periodically,
in times $t = [1, \ldots, T]$. 
  The technical question of \emph{how} the adversary can count the stack of ballots is discussed in Section~\ref{section:counting};
  we only mention that one might use, for example, accurate weight scales, laser based measurement equipment, or banknote counters.
The adversary also collects the time of vote for each voter.
This information is already collected by the local elections committee,
and is sent to the parties via a dedicated form called ``Tofes-1000'' (which translates to ``1000-Form'').

We define a \emph{frame} to be the time period between two consecutive counts.
Through their voting times,
we can divide the voters into frames,
and assign a probability distribution to their vote according to the count of the respective frame.
We refer to the set of voters
between the count in time $t - 1$ and the count in $t$,
in election cycle $u$,
as $V^{u,t}$,
and refer to the probability distribution associated with this time frame as $C^{u,t}$.
Notice that we have $t$ frames: frame $1$ to frame $t$.
The probability distribution $C^{u,t}$ can be represented as a vector,
such that each element in it corresponds to a party~$p$,
and each value in it is equal to the number of ballots of party $p$ which are missing from the stack in this frame,
normalized by the total number of voters in the frame.
For example, if in the second elections
Alice voted for the party named \emph{Meretz} between time $t = 5$ and time $t = 6$,
then we have that the set $V^{2,5}$ contains Alice and $C^{2,5}[Meretz] > 0$.
It follows that, initially, the size of the anonymity set of every voter $v \in V^{u,t}$
is at most the number of non-zero items in $C^{u,t}$ (and not the number of non-zero items in the tally of the whole polling station).

Notice that
using these frames,
the adversary can recreate the real tally of each polling station.
However,
the adversary can also directly collect the real tally of each polling station,
since this information is published by the central elections

Indeed,
from the perspective of each voter,
every election cycle is composed of exactly one frame to which she belongs
and an arbitrary number of frames to which she does not belong.

\subsection{The Attack Algorithm}
\label{sec:EBN}

Our algorithm is composed of the following three functions.
\begin{itemize}
	\item The \textbf{\homo{}} function iterates over all frames, searching for homogeneous ones, i.e., frames in which all voters voted for the same party. If such a frame is found, then all voters in it are assigned to this party, the size of the frame is subtracted from the tally of that party, and the voters in the frame are removed from all other frames they participate in. 
	\item The \textbf{\single{}} function iterates over all voters. For each voter, it intersects the frames in which it participates,
	to find which parties are shared by all involved frames. If only a single party is shared between all frames in which a voter participates, then it assigns this party to the voter. The tally for this party is then reduced by 1
and the corresponding frame counts are updated. 
	\item The \textbf{\LE{}} function iterates over tuples of (voter, party, frame).
	For each such tuple,
	it estimates,
	independently for each frame,
	the likelihood that a voter in the frame voted for each of the parties involved in that frame. The likelihood is calculated as the number of votes which the party got in this frame over the number of voters in this frame. The likelihood for a voter to vote for a certain party is the product of the respective probabilities in all frames she participated in. The output of this function is a matrix $L$ where each row $v$ is a voter, and each column $p$ is a party. An element $L_{v,p}$ in this table is the likelihood that a voter $v$ voted for a party $p$. We search for the pair $(v,p)$ giving the largest value $L_{v,p}$ and assign the voter $v$ to the party $p$. The tally is then decreased by one for that party $p$ and the corresponding frame counts are updated.
\end{itemize}	

The attack algorithm is composed of two phases: the \emph{\firstphase{}} and the \emph{\secondphase{}}. In the \firstphase{} we call \homo{} and \linebreak
\single{} over and over until no new assignments can be made. This phase is safe in the sense that whenever the algorithm assigns a party to a voter, this assignment is necessarily correct. In other words, it can either return the right party for a voter, or output a symbol indicating that it was unable to de-anonymize her.
In Section \ref{sec:veal}, we present the success rate of the algorithm when only this phase is being used. 

In the \secondphase,
which we invoke after no more voters can be de-anonymized through the \firstphase,
the \LE procedure is used for making a probabilistic decision,
assigning a party to a single voter for which we are most certain about.
We then start over the process of calling to Find\_ \linebreak Homogeneous\_Frames and \single until they can no longer de-anonymize voters,
in which case we call \LE again.
The algorithm halts when all voters have been assigned to parties.
Note that
during the course of this phase,
\homo and \single can err due to previous wrong guesses made by \LE. However, as we will see in Section \ref{sec:veal}, although the \secondphase can make wrong guesses, its success probability is much higher than that of \firstphase, suggesting that it usually does not. 
A pseudocode of the attack is given in Algorithm~\ref{pseudocode}.

\renewcommand{\algorithmicrequire}{\textbf{Input:}}
\renewcommand{\algorithmicensure}{\textbf{Output:}}
\newcommand{\INDENT}[1][1]{\hspace{#1\algorithmicindent}}
\newcommand{\MYCOMMENT}[1]{\{\textbf{#1}\}}

\begin{algorithm*}[p]
\caption{Pseudocode of the attack for a certain polling station.}
\label{pseudocode}
\begin{algorithmic}
\REQUIRE List of voters $V^{u,t}$ for $t \in [T]$ and $u \in [U]$ (list of voters)
\REQUIRE Normalized frame counts $C^{u,t}$ for $t \in [T]$ and $u \in [U]$ (one value per party; sums to $1$)


\STATE
\MYCOMMENT{\Firstphase{}}
\WHILE{progress is made}

\STATE
\MYCOMMENT{\homo{}}
  \FOR{$u \in [U]$; $t \in [T]$; party $p$}
    \IF{$C^{u,t}[p] = 1$ (and thus, for each $p' \neq p$, we have $C^{u,t}[p'] = 0$)}
      \STATE assign all voters in $V^{u,t}$ to $p$ and decrease the tally of $p$ by $|V^{u,t}|$
    \ENDIF
  \ENDFOR

  \STATE
\MYCOMMENT{\single{}}
  \FOR{voter $v$}
    \IF{$\cap_{u \in [U], t \in [T], v \in V^{u,t}} \{p : C^{u,t} > 0\} = \{p\}$}
      \STATE assign $v$ to $p$, decrease the tally for $p$ by one, and update $C^{u,t}$
    \ENDIF
  \ENDFOR

\ENDWHILE

\STATE
\MYCOMMENT{\Secondphase{}}

\WHILE{not all votes have been extracted}

\STATE
\MYCOMMENT{\LE{}}
\FOR{voter $v$; party $p$}
  \STATE compute likelihood of $v$ voting for $p$ as $L^{v,p} = \Pi_{u \in [U], t \in [T], v \in V^{u,t}} {C^{u,t}[p]}$  
\ENDFOR

\STATE let $v'$ and $p'$ be the pair for which the likelihood value $L^{v',p'}$ is maximal
\STATE assign $v'$ to $p'$, decrease the tally for $p$ by one, and update $C^{u,t}$

 \WHILE{progress is made}

  \MYCOMMENT{\homo{}}
  \FOR{$u \in [U]$; $t \in [T]$; party $p$}
    \IF{$C^{u,t}[p] = 1$ (and thus, for each $p' \neq p$, we have $C^{u,t}[p'] = 0$)}
      \STATE assign all voters in $V^{u,t}$ to $p$ and decrease the tally of $p$ by $|V^{u,t}|$
    \ENDIF
  \ENDFOR
  
  \MYCOMMENT{\single{}}
  \FOR{voter $v$}
    \IF{$\cap_{u \in [U], t \in [T], v \in V^{u,t}} \{p : C^{u,t} > 0\} = \{p\}$}
      \STATE assign $v$ to $p$, decrease the tally for $p$ by one, and update $C^{u,t}$
    \ENDIF
  \ENDFOR

\ENDWHILE

\ENDWHILE
\end{algorithmic}
\end{algorithm*}

\section{Evaluation of the Attack}
\label{sec:veal}

In this section we evaluate, through simulations, the success rate of the attack proposed in Section~\ref{sec:attack}.
The model considered here assumes that voters do not change their minds between election cycles. We defer the justification of this assumption to Section~\ref{sec:discussion}.
We also assume, for the sake of simplicity, that voters always vote in the same polling station, and that no new voters join or leave the registry.

\subsection{Simulations}
\label{ssec:vta}

To calculate the success rate of the attack,
we ran simulations based on the results of the 2013 general elections in Israel
as published by the general elections committee~\cite{voting_results}.
In these elections,
Israel's eligible voters were divided into 9879 polling stations.
The law upper-bounds the maximal number of eligible voters assigned to a polling station at 900;
in practice,
the maximal number of voters assigned to a polling station was 894,
and the median number of voters assigned to each polling station was 590.
The voting turnout was low, and out of the 5,654,842 eligible voters only 3,617,857 (64\%) actually voted;
as a result,
the median number of actual voters per polling station was 366.
Out of these,
a total number of 3,579,793 votes were counted as legitimate votes.\footnote{%
  Absentee votes
  (that is, voters who do not vote in their assigned ballot, such as diplomats, soldiers, and seamen),
  which account to about 5\% of the votes,
  are excluded for simplicity.}

We model each polling station independently of all other polling stations,
as we see no dependencies between different polling stations.\footnote{This independence implies that an adversary can focus their effort on subsets of polling stations which are of interest, or where they expect to achieve a high success rate.}
The published results include,
per polling station,
the number of assigned voters,
the number of voters who arrived,
the number of legitimate votes,
the number of votes received by each party per polling station,
and an accumulated turnout rate per two hours. 

Due to obvious reasons we do not have the real data needed to actually run the attack,
although we do use real data from the tallies of the various polling stations.
We therefore resort to the ``second-best'' option and use a simulation of the elections process.
We denote the number of voters in the attacked polling station by $n$
and set the number of frames $T$ to be either $30$, $15$, or $7$:
  for the vast majority of the polling stations,
  this corresponds to counting the ballots once in half an hour, an hour, or two hours.\footnote{When $T=7$ the first count is done after 3 hours.}
We created $n$ ``virtual'' voters, and split them randomly over the frames according to the turnout rate.
For each frame we ``counted'' the number of missing ballots,
and built the voting distribution for it.
This procedure is repeated $U$ times,
corresponding to $U$ consecutive election cycles;
we chose $U = \{2, 3, 4, 5, 6, 7\}$.

\subsection{Results}
\label{ssec:results}

We begin by reporting and analyzing our results,
where we set $T$ to be $30$. 
Later we report on simulations done with $T = 15$ and $T = 7$.

\medskip\noindent\textbf{Average success rate}
The average success rate of the attack (over the polling stations)
is provided in Table~\ref{table:resultsparty}.
The baseline is the success rate had the adversary always assigned the largest party or political group to all voters of the ballot box.

\begin{table}[t]
\centering
\caption{Average success rate of the attack, for $T = 30$, for extracting the exact party that the voters voted for, and the political group that the voters belong to.
The baseline is $38\%$ for extracting the party and $54\%$ for extracting the group.}
\vspace{5px}
\begin{tabular}{c@{\hskip 5px}|@{\hskip 5px}c@{\hskip 15px}c@{\hskip 15px}c}
Election cycles & \firstphase & \secondphase, party & \secondphase, group \\ \hline
$2$ & $ 7\%$ & $46\%$ & $59\%$ \\
$3$ & $19\%$ & $63\%$ & $73\%$ \\
$4$ & $35\%$ & $76\%$ & $83\%$ \\
$5$ & $50\%$ & $84\%$ & $89\%$ \\
$6$ & $62\%$ & $90\%$ & $93\%$ \\
$7$ & $71\%$ & $93\%$ & $96\%$
\end{tabular}
\label{table:resultsparty}
\end{table}

When trying to recover the political group that a voter voted for we first let the algorithm assign a party to the voter and count it as a success if this party is part of the correct group. Since the \firstphase cannot output incorrect assignments, the success rates do not change for that phase. In contrast, we can see in the table that for the \secondphase, the success rate increases in all cases. 

The more natural course, where we first merge the parties into political groups and then run the algorithm with 6 ``virtual'' parties, was tried but offered inferior results compared with the selected approach. Consider the following scenario: a voter $v_1$ voted for party 1 and shares a frame in $u=2$ with a voter $v_2$ who voted for party 2 and in $u=3$ with a voter $v_3$ who voted for party 3. Assume that parties 2 and 3 are of the same political group. Now, before merging them we could exclude parties 2 and 3 as possible parties for $v_1$. This is no longer possible after the merge as $v_2$ and $v_3$ are indistinguishable. 

\medskip\noindent\textbf{Size and homogeneity}
For a more detailed understanding of the factors which affect our success rate,
we provide further results.
Specifically,
We show the success rate of the attack as a function of the polling station size,
and the homogeneity of the polling station
(the homogeneity of a polling station is defined to be the standard deviation of its normalized tally with respect to the unanimous vector, i.e., the squared root of the squared difference between the frame and a frame where all parties got the same number of votes, normalized by the number of voters),
both for the \firstphase of the algorithm and for the \secondphase of the algorithm,
for $U = \{2, 3, 4, 5, 6, 7\}$ election cycles.

Further,
we consider the attack as trying to reveal either (1) the exact party for which the voters voted for,
or (2) the political group for which the voters voted for.
Specifically,
the political parties in Israel, as of 2013, can be grouped into six almost distinct groups:
  left (Meretz and HaAvoda),
  right (Habait Hayehudi, Likud, and Otzma Leisrael),
  center\footnote{Sometimes referred to as ``secular''.} (Eretz Chadasha, Kadima, Or, Yesh-Atid, and Hatnuaa),
  ultra orthodox Jews (Yahadut Hatora, Am Shalem, and Shas),
  Arabs (Balad, Hatikva-Leshinui, Chadash, Raam, and Daam), 
  and MISC (all the other parties, all of which do not meet the election threshold for representation).

The corresponding figures are given in Figures
\ref{figure:safe_success_by_size_for_party},
\ref{figure:unsafe_success_by_size_for_party},
and
\ref{figure:unsafe_success_by_size_for_gush}.
In those graphs,
we show results for $U = \{2, 3, 4, 5\}$,
and do not visualize the results for $U = \{6, 7\}$,
to not clutter the image too much,
and since the point is already clear with those values.

\resultssubfigure{safe}{size}{party}{party}{size}{safe}{entropy}{party}{party}
\resultssubfigure{unsafe}{size}{party}{party}{size}{unsafe}{entropy}{party}{party}
\resultssubfigure{unsafe}{size}{gush}{political group}{size}{unsafe}{entropy}{gush}{political group}

\subsubsection{Results Analysis}

There are several important variables which affect our success rate.
First,
as one might expect,
using more election cycles (that is, increasing $U$),
or aiming at finding only the political group for which the voters voted for,
increases the success rate of the algorithm.
Second,
the \secondphase indeed increases the success rate of the attack,
however at the cost of sometimes making wrong decisions and assigning wrong parties to some voters.

The other two important variables are the size of the polling station and the homogeneity of the polling station.
Specifically,
it is apparent that 
the strongest factor on our success rate is the size of the polling station.
Indeed, we see that the polling station's size and the success rate are highly correlated;
concretely,
the smaller the polling station is,
the higher the success rate. 

Less strong than the size of the polling station,
the homogeneity of the polling station is an important factor on the success rate of the algorithm.
(Recall that we measure the homogeneity of a polling station as the standard deviation of its normalized tally.)
Specifically,
it seems that the more homogeneous the polling station is,
the better the attack performs.
Interestingly,
the correlation is decreasing as we consider more election cycles. 

The opposing trends of these correlations suggest that, as the number of considered election cycles grow, the importance of the homogeneity decreases in favor of the size of the polling station which becomes more prominent. 

For validation,
the Pearson correlation between the polling station's size and the success rate, and the polling station's homogeneity and the success rate, are given in Tables~\ref{table:correlationsizesuccessrate} and \ref{table:correlationsizehomogeneity} when considering the \firstphase, the \secondphase when the exact party is extracted, and the \secondphase when the political group is extracted. 

\begin{table}[t]
\centering
\caption{Pearson correlation between the polling station's size and homogeneity to the success rate for extracting the exact party of voters, using the \firstphase.}
\vspace{5px}
\begin{tabular}{c|c|c}
& size & homogeneity \\ \hline
$2$ election cycles & $-0.56$ & $0.29$ \\
$3$ election cycles & $-0.70$ & $0.17 $ \\
$4$ election cycles & $-0.76$ & $0.09$ \\
$5$ election cycles & $-0.76$ & $0.01$ \\
\end{tabular}
\label{table:correlationsizesuccessrate}
\end{table}

\begin{table}[t]
\centering
\caption{Pearson correlation between the polling station's size and homogeneity to the success rate for extracting the exact party of voters
and the political group, using the \secondphase.
Each cell contains two numbers,
the first of which corresponds to the exact party while the second corresponds to the political group.}
\begin{tabular}{c|c|c}
& size & homogeneity \\ \hline
$2$ election cycles & $-0.64$, $-0.36$ & $0.57$, $0.81$ \\
$3$ election cycles & $-0.81$, $-0.62$ & $0.30$, $0.56$ \\
$4$ election cycles & $-0.83$, $-0.70$ & $0.16$, $0.38$ \\
$5$ election cycles & $-0.80$, $-0.70$ & $0.05$, $0.27$ \\
\end{tabular}
\label{table:correlationsizehomogeneity}
\end{table}

Importantly,
the size of the polling station seems to be not correlated with its homogeneity
(in fact, the Pearson correlation between these two variables is as low as 0.04).

\subsection{Further Experiments}\label{section:further-experiments}

In this section,
we report on results of our simulations with varying interval times for counting the ballots.
Specifically,
the results from the previous section were for $T = 30$,
corresponding
(for almost all polling stations)
to counting the ballots once in half an hour.
Next,
in Table~\ref{table:resultsseven},
we report the average success rate of the attack (over the polling stations)
for $T = 15$ and $T = 7$,
corresponding
(for almost all polling stations)
to counting the ballots once in an hour
and once in two hours.

\begin{table}[t]
\centering
\caption{Average success rate of the attack, for extracting the exact party that the voters voted for, and the political group that the voters belong to,
for $T = 15$ and $T = 7$,
that is,
when counting $15$ times a day and $7$ times a day.
}
\vspace{5px}
\begin{tabular}{c@{\hskip 5px}|@{\hskip 5px}c@{\hskip 15px}c@{\hskip 15px}c}
Election cycles & \firstphase & \secondphase, party & \secondphase, group \\ 
		& $T = 15,T=7$& $T = 15,T=7$& $T = 15,T=7$\\	\hline
$3$ & $ 3\%$, $0.6\%$ & $41\%$, $30\%$ & $55\%$, $46\%$ \\
$4$ & $ 5\%$, $0.9\%$ & $47\%$, $33\%$ & $60\%$, $49\%$ \\
$5$ & $ 7\%$, $1.2\%$ & $53\%$, $36\%$ & $65\%$, $51\%$ \\
$6$ & $ 9\%$, $1.4\%$ & $59\%$, $38\%$ & $69\%$, $53\%$ \\
$7$ & $12\%$, $1.6\%$ & $63\%$, $41\%$ & $72\%$, $54\%$
\end{tabular}
\label{table:resultsseven}
\end{table}  



\section{Discussion}
\label{sec:discussion}

In this section,
we begin by briefly discussing various methods for counting the ballots and the time intervals by which an adversary is able to perform those counts. We continue by discussing some consequences of our research. Then, we discuss countermeasures which can be taken in order to guard the system against attacks
as the one described in this paper. Finally, we discuss possible ways of extending our attack when we allow voters to change their minds between election cycles.

\subsection{Counting the Ballots}
\label{section:counting}

The question of how exactly to count the ballots is somehow beyond the scope of the current paper,
however,
we do mention some methods bellow,
which seem to be sufficient for our needs.
As examples,
one might use accurate weight scales;
one might use laser-based measurement equipment;
or one might use banknote counters.

Notice that, during election day,
members of the polling station committee are allowed,
and encouraged,
to go behind the curtains once in a while to check that all parties have sufficient ballots.

We remark also that there is no need for a nation-wide systematic attack,
as the polling stations are independent of each other,
and it is sufficient to perform the attack on each polling station on its own, thus allowing to focus the efforts on high priority polling stations. 

\subsection{Putting the Results in Context}
We now give examples for countries where a similar voting system is being used and discuss possible consequences in their context.

Our first example is Algeria \cite{Algeria} where the young democracy is still struggling with conducting free elections. During the elections there have been numerous reports about voting-related violence and it is not unreasonable to believe that voting for the ``wrong'' candidate may put someone under physical danger. 

Even in less extreme cases such as Israel, there may be unwanted repercussions such as government-led investments made to prefer some voters over others. This has been more prominent in the early years of Israel, where better rations where given to members of Mapai, the ruling party at the time. Such blunt favoritism has been long abolished now but even today the phenomenon of voting-contractors still exists; a voting-contractor is a person having the power to tell a large group of people how to vote. The power of a voting-contractor is determined by the number of people they can enlist. It is very hard nowadays for a party to contest without soliciting such voting-contractors and this activity is not even being conducted in secret anymore. 

Finally, even in countries where the government is unlikely to act dubiously such as Sweden \cite{Sweden} there may still be social consequences for not voting as everybody else in the village.
Finally,
we mention Spain~\cite{Spain} and France~\cite{France} as two further countries where similar voting systems are used.   

\subsection{Countermeasures}
\label{sec:counter}

In this section, we briefly present possible countermeasures for the attack. The most obvious countermeasure is switching to cryptographically secure voting systems.
Such systems are not only better understood than traditional ones, but they also allow to quantify the security loss in various scenarios. 

Should a paper based system is still desired, we note that the weakness of the system comes from the fact that the stack of tickets available to the voters
``remembers''
all previous choices.
This weakness can be avoided by changing the ballot to a one
that requires the voter to choose an item from a closed list printed beforehand;
consider, for example, the ballots used in most countries of the EU. 
An additional advantage of such a ballot is that it allows the voters to rely a more complex decision
(for example,
reordering the members of the list as done in Europe,
or moving the vote to another party as done in Australia).

Another improvement that can be introduced into the system is to not allow any information to leave the polling station.
The current law in Israel already disallows any form of radio communication.
Extending the law to prevent any transfer of information but the tally outside the polling station
(both during and after the elections),
would make processing such information illegal for third parties, moving our attack from the ``gray area'' to the black.

Finally, as the obligation to conduct fair elections is the role of the government, it may be useful to develop a mathematical model that will take both heterogeneity and polling stations' sizes to help decision makers to reassign voters to voting precincts. 

\subsection{Allowing Voters to Switch Parties Between Election Cycles}
\label{ssec:JAH}

The whole purpose of holding elections is to allow people to change the composition of the governing bodies.
The reason we assume that voters do not change their behavior is made for the sake of simplicity.
We can loosen this restriction completely and allow each voter to choose the party she votes for in every election cycle,
even uniformly at random.
This would be,
however,
too extreme,
since most voters do not tend to change their viewpoints dramatically between election cycles.

Intuitively,
in a multi-partied system,
a voter who voted for party $p$ in one elections cycle will probably vote for a party ideologically close to $p$ in the successive cycles.
There is actually some concrete evidence supporting the above intuition,
as we discuss next.

Indeed,
by analyzing election surveys provided by the \emph{Israel National Election Studies}~\cite{2009,2013},
we found out that roughly 50\% of the voters did not change their vote between the 2009 and the 2013 elections
(this number becomes roughly 60\% if we count the successor of a party as not necessarily the one which inherited its name,
but the one which is ideologically closest.\footnote{%
  Due to the somewhat unstable political system of Israel,
  a large amount of people cannot find their political home in any of the existing parties,
  and tend to vote in every elections cycle to a newly ``trending'' party.
  Moreover,
  parties often split, merge, or change their names.}.

Moreover, when groups of parties are being considered, the change is insignificant. In fact, the change in the political map between the 2015 and the 2013 elections was that only a one seat (corresponding to 0.83\% of the elected seats) moved between the groups. 
These numbers mean that we can simply run our attack without accounting for voters which change their minds,
and we expect to preserve a fairly high success rate.
Moreover,
one could take such information into account;
we next discuss one possible way of doing so.

In our attack,
instead of computing the likelihood of each voter to vote for a specific party in all election cycles,
we can compute the likelihood of each voter to vote for a list of different parties (one element per each election cycle);
then,
given the information encoded in the transition matrices,
we can multiply each likelihood by the `global` likelihood of such a vote.

We were not able to perform simulations for such scenarios
since we do not have the real votes of voters across election cycles.
That is,
while we have the tallies for each election cycle,
we can not infer the real turnover, i.e., which votes correspond to which voters in different election systems.

\section{Conclusion and Future Work}
\label{sec:conc}

Free elections are an essential element in modern liberal democracies.
In this paper,
we presented a way to attack the Israeli voting system
(as well as several other similar systems),
showing that it is possible to recover the votes of voters in this system.
Specifically,
this is possible using a very small amount of additional public information,
which includes the results of the elections,
the time of vote per voter,
and a periodical count of the ballots from the tray. 

We would like to end with some ideas for future research and extensions of this attack.
First,
since the attack assigns voters to the parties they voted for,
it sounds reasonable that,
using flow techniques
(which are successfully being used for assignment problems),
we might improve the success rate of the attack.
Second,
since the \firstphase of the attack is based on evaluating constraints on the possible parties
for which each voter might have voted for,
it sounds reasonable that using constraint satisfaction techniques might improve the success rate of the attack.

\subsubsection*{Acknowledgment}
The authors would like to thank Aviad Stier for bringing into our attention the fact that the parties collect the time of vote of all voters, Adv. Jonathan J. Klinger for assisting us with our petition to the Israeli general elections committee, Amihai Bannett for permitting us to use hit photos in this paper and in a poster that presented preliminary results of this work \cite{DBLP:conf/ccs/AshurD13}. We thank Dubi Kanengisser for pointing us to good resources in the field of political studies.  Special thanks are to colleagues with which the authors discussed this research, specifically to Gustavo Mesch, Claudia Diaz, Tamar Zondiner, Yair Goldberg, Atul Luykx, Alan Szepieniec, Shir Peled, as well as to the anonymous referees.

The first author was partially supported by the Research Fund KU Leuven, OT/13/071 and by European Union’s Horizon 2020 research and innovation programme under grant agreement No 644052 HECTOR and grant agreement No H2020-MSCA-ITN-2014-643161 ECRYPT-NET. The second author was supported
in part by the Israeli Science Foundation through grant No. 827/12 and by the
Commission of the European Communities through the Horizon 2020 program
under project number 645622 PQCRYPTO. The third author is supported by a postdoctoral fellowship from I-CORE ALGO.

\bibliographystyle{plain}
\bibliography{bib}

\end{document}